\documentclass[twocolumn,showpacs,preprintnumbers,amsmath,amssymb,prb,aps]{revtex4-1}
\usepackage{epsfig}
\usepackage{epstopdf}
\usepackage{graphicx}
\usepackage{dcolumn}
\usepackage{bm}
\usepackage{slashbox}
\usepackage{textcomp}
\begin{document}

\title{Electronic structure study of vanadium spinels by using density functional theory and dynamical mean field theory}
\author{Sohan Lal}
\altaffiliation{Electronic mail: goluthakur2007@gmail.com}
\author{Sudhir K. Pandey}
\affiliation{School of Engineering, Indian Institute of
Technology Mandi, Kamand - 175005, India}

\date{\today}

\begin{abstract}
 
     Theoretically, various physical properties of AV$_{2}$O$_{4}$ (A=Zn, Cd and Mg) spinels have been extensively studied for last 15 years. Besides of this, no systematic comparative study has been done for these compounds, where the material specific parameters are used. Here, we report the comparative electronic behaviour of these spinels by using a combination of density functional theory and dynamical mean-field theory, where the self-consistent calculated Coulomb interaction $U$ and Hund's coupling $J$ (determined by Yukawa screening $\lambda$) are used. The main features, such as insulating band gaps ($E_{g}$), degree of itinerancy of V 3$d$ electrons and position of lower Hubbard band are observed for these parameters in these spinels. The calculated values of $E_{g}$ for ZnV$_{2}$O$_{4}$, CdV$_{2}$O$_{4}$ and MgV$_{2}$O$_{4}$ are found to be $\sim$0.9 eV, $\sim$0.95 eV and $\sim$1.15 eV, respectively, where the values of $E_{g}$ are close to experiment for ZnV$_{2}$O$_{4}$ and MgV$_{2}$O$_{4}$. The position of lower Hubbard band are observed around $\sim$-1.05 eV, $\sim$-1.25 eV and $\sim$-1.15 eV for ZnV$_{2}$O$_{4}$, CdV$_{2}$O$_{4}$ and MgV$_{2}$O$_{4}$, respectively, which are also in good agreement with the experimental data for ZnV$_{2}$O$_{4}$. The order of average impurity hybridization function of V site are found to be ZnV$_{2}$O$_{4}$$>$MgV$_{2}$O$_{4}$$>$CdV$_{2}$O$_{4}$. Hence, the degree of localization of V 3$d$ electrons is largest for CdV$_{2}$O$_{4}$ and smallest for ZnV$_{2}$O$_{4}$, which is in accordance with our earlier results. Hence, present work shows the importance of material specific parameters to understand the comparative electronic behaviour of these compounds.

\end{abstract}

\pacs{71.27.+a, 71.20.-b, 71.15.Mb}

\maketitle

\section{Introduction} 

  Strongly correlated vanadium spinels, AV$_{2}$O$_{4}$ (A=Zn, Cd and Mg) have been thoroughly studied both theoretically and experimentally because of the presence of large variety of physical properties such as electronic, structural and magnetic etc.\cite{Tsunetsugu,Tchernyshyov,Matteo,Motome,Maitra,Pandey2011,Lal1,Lal3,Lal5,Lee,Suzuki,Nishiguchi,Wheeler,Reehuis} At a room temperature, these compounds shows face-centered-cubic (FCC) structure. Geometrical frustration is seen in the antiferromagnetically coupled V ions with spin $S$=1 reside on the vertices of corner-sharing tetrahedra forming a pyrochlore lattice.\cite{Tchernyshyov,Tsunetsugu,Lee,Matteo,Suzuki} In these system, the interplay of spin, electron, orbital and lattice degrees of freedom leads to two successive phase transitions. The structural and magnetic transition takes place from cubic to tetragonal and paramagnetic to anti-ferromagnetic, respectively.\cite{Nishiguchi,Reehuis,Wheeler} The structural (magnetic) transition temperature {\it $T_S$} ({\it $T_N$}) for ZnV$_{2}$O$_{4}$, MgV$_{2}$O$_{4}$, and CdV$_{2}$O$_{4}$ are $\sim$50 K ($\sim$40 K), $\sim$65 K ($\sim$42 K) and $\sim$97 K ($\sim$35 K).\cite{Nishiguchi,Reehuis,Wheeler,Takagi} However, their Curie-Weiss temperature are found to be much higher than the room temperature.\cite{Takagi} {\it $T_S$}$>${\it $T_N$}, indicates that the certain degree of geometrical frustration is still present in these systems. The degree of geometrical frustration is measured by the frustration index, which is the ratio of Curie-Weiss temperature and {\it $T_N$}. The frustration index is largest for ZnV$_{2}$O$_{4}$ (21.3) and lowest for CdV$_{2}$O$_{4}$ (11.4) as compared to MgV$_{2}$O$_{4}$ (14.3).\cite{Takagi} The experimentally observed magnetic moment per V atom (band gap) for ZnV$_{2}$O$_{4}$, MgV$_{2}$O$_{4}$ and CdV$_{2}$O$_{4}$ are $\sim$0.63 $\mu$$_{B}$ ($\sim$0.32 eV, $\sim$1.1 eV), $\sim$0.47 $\mu$$_{B}$ ($\sim$0.36 eV, $\sim$1.08 eV) and $\sim$1.19 $\mu$$_{B}$ ($\sim$3.16 eV), respectively.\cite{Nishiguchi,Reehuis,Wheeler,Pardo,Rogers} Normally the degree of localization of electrons is measured by the ratio of transfer integral ($t$) between neighbouring site and Coulomb interaction $U$. Canosa $et$ $al$. have argued qualitatively that the order of degree of localization of electrons of V site for CdV$_{2}$O$_{4}$$>$ZnV$_{2}$O$_{4}$$>$MgV$_{2}$O$_{4}$.\cite{Canosa} While we have shown quantitatively in earlier paper that the degree of localization of electrons of V site is largest for CdV$_{2}$O$_{4}$ and smallest for ZnV$_{2}$O$_{4}$.\cite{Lalarxiv} 
    
    In order to understand the various physical properties as mentioned above for these compounds, several groups have proposed different mechanisms.\cite{Tsunetsugu,Tchernyshyov,Matteo,Motome,Maitra,Pandey2011,Lal1,Lal3,Lee,Yamashita,Khomskii} Most of these mechanisms are based on the model calculations, which are fully parameters dependent. Similarly, some density functional theory (DFT)+$U$ electronic structure calculations have also been performed in these systems, where normally $U$ and $J$ are used as adjustable parameters. By adjusting these parameters, different groups have explained the various experimentally observed properties in these systems.\cite{Maitra,Pandey2011,Lal1,Kaur} There is no systemically comparative studied of these compounds available in the literature, where the parameters, $U$ and $J$ are material specific. Also, the limitations of this method has been clearly observed, if one want to understand the spectral properties of strongly correlated systems. This is because of its naturally static treatment of electronic correlation. For example, this method fails to reproduce the experimentally reported quasiparticle peak of 5$f$ character of $\delta$-Pu near the Fermi level.\cite{Zhu} Hence, in order to reproduce the spectral and thermodynamic properties of strongly correlated materials, a more sophisticated methods are needed to account the correlation effects beyond this method. The most successful method is the dynamical-mean-field theory (DMFT), which describe the correlation effects in a periodic lattice by a strongly interacting impurity coupled to a self consistent bath.\cite{Georges1992,Georges1996} Merging this with DFT method, DFT+DMFT approach provides more material specific predictions of correlation effects in solids.\cite{Anisimov, Kotliar,Martins,Biermann,Roekeghem} Despite of the many success of this method, Wang $et$ $al$. and Dang $et$ $al$. have shown that the Mott insulating state in early and late transition metal oxides are described by tuning of several parameters, including double counting (DC) and interaction $U$.\cite{Wang,Dang} However, recently Haule $et$ $al$. have developed a method, where no fine tuning of parameter is required to predict the Mott gaps in early transition metal oxides. They have shown that for fixed value of $U$, Mott gaps of order of experimental data are observed for these compounds.\cite{Haule} 
    
    From above discussion, it is clear that none of the group have applied DFT+DMFT method in the strongly correlated vanadium spinels even having the limitations of DFT+$U$ method as mentioned above. Here, in present work we try to understand the comparative electronic behaviour of strongly correlated vanadium spinels by using DFT+DMFT method, where the self-consistently computed material specific parameters are used. For ZnV$_{2}$O$_{4}$ and MgV$_{2}$O$_{4}$, the calculated values of $E_{g}$ are in good agreement with the experimental data. Among these spinels, the average impurity hybridization function of V site is largest for ZnV$_{2}$O$_{4}$ and smallest for CdV$_{2}$O$_{4}$. Hence the order of degree of localization is found to be CdV$_{2}$O$_{4}$$>$MgV$_{2}$O$_{4}$$>$ZnV$_{2}$O$_{4}$, which is consistent with our earlier reported data. The position of lower Hubbard band observed in the present study is also consistent with the experimental data for ZnV$_{2}$O$_{4}$.
  
\section{Computational details}
   
    The paramagnetic electronic structure calculations of vanadium spinels carried out in FCC phase have been divided into two parts, DFT and DFT+DMFT. DFT part of the calculations are preformed by using the full-potential linearized augmented plane-wave (FP-LAPW) method as implemented in WIEN2K code.\cite{Blaha} The exchange-correlation functional has been treated within the GGA of PBEsol.\cite{Perdew} The experimentally observed crystal structure for these spinels are taken from the literature.\cite{Onoda,Reehuis,Wheeler} The muffin-tin sphere radii used for Zn, Cd, Mg, V and O atoms are 2.0, 2.46, 1.39, 2.0 and 1.54 Bohr, respectively. k-point grid size of 2000 points in the whole Brillouin zone have been used here. Every calculations are converged with charge convergence per cell below 10$^{-4}$ electronic charge. In order to described the itinerant and localized behaviours of correlated electrons on equal footing, we have used DFT+DMFT method as implemented in the study by Haule $et$ $al$.\cite{Haule2010} DFT+DMFT calculations are carried out at room temperature, which are fully self-consistent in the electronic charge density and impurity levels. Continuous time quantum Monte Carlo impurity solver has been used to solve the auxiliary impurity problem.\cite{Haule2007} The exact DC scheme proposed by Haule has been used here.\cite{Haule2015} V t$_{2g}$ shell are treated within DMFT. Density-density form of the Coulomb repulsion has been used in all calculations. The values of $U$ used here are taken from our earlier paper, where the self-consistently calculated values of $U$ for ZnV$_{2}$O$_{4}$, CdV$_{2}$O$_{4}$ and MgV$_{2}$O$_{4}$ are 5.9, 5.7 and 6.2 eV, respectively.\cite{Lalarxiv} Using these values of $U$, the Yukawa screening $\lambda$ (=1.34, 1.38 and 1.28 a.u.$^{-1}$ for ZnV$_{2}$O$_{4}$, CdV$_{2}$O$_{4}$ and MgV${2}$O$_{4}$, respectively) were uniquely determined through the matrix elements of the Yukawa interaction in DMFT basis. Similarly, the values of Hund's coupling $J$ (=1.0, 0.99 and 1.02 eV for ZnV$_{2}$O$_{4}$, CdV$_{2}$O$_{4}$ and MgV${2}$O$_{4}$, respectively) were uniquely determined by $\lambda$ through Yukawa form of Coulomb interaction.\cite{Haule2015} Hence, all the parameters ($U$, $\lambda$ and $J$) used in the present study were material specific. The convergence of the charge/cell of these systems are set to be less than 10$^{-4}$ electronic charge. All these calculations are converged on the imaginary axis. Next, to obtain the self energy on the real axis, we need to do analytical continuation. The maximum entropy method is used for analytical continuation of the self energy from the imaginary frequency axis to real frequencies to obtain spectra on the real axis.\cite{Jarrell}       
 
\section{Results and Discussion}

\begin{figure}[h]
  \begin{center}
    \includegraphics[width=0.45\textwidth]{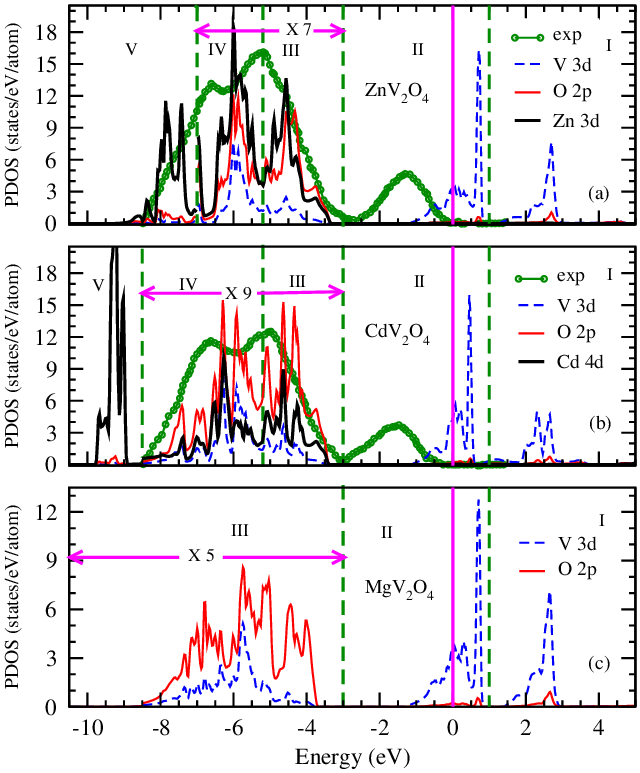}
    \label{}
    \caption{(Color online) Partial density of states (PDOS) of (a) Zn 3$d$, V 3$d$ and O 2$p$ states for ZnV$_{2}$O$_{4}$, (b) Cd 4$d$, V 3$d$ and O 2$p$ states for CdV$_{2}$O$_{4}$ and (c) V 3$d$ and O 2$p$ states for MgV$_{2}$O$_{4}$ compounds computed within DFT. For better comparison, PDOS in regions III and IV are multiplied by a factor of seven (nine) for ZnV$_{2}$O$_{4}$ (CdV$_{2}$O$_{4}$), while in region III, it is multiplied by a factor of five for MgV$_{2}$O$_{4}$ compound. Zero energy corresponds to the Fermi level. Integrated background subtracted X-ray photoemission spectroscopy measurements data for ZnV$_{2}$O$_{4}$ and CdV$_{2}$O$_{4}$ are taken from the Ref. 41.}
  \vspace{-.8cm}
  \end{center}
\end{figure}
  
    The plot of partial density of states (PDOS) of Zn 3$d$, Cd 4$d$, V 3$d$ and O 2$p$ states of AV$_{2}$O$_{4}$ (A=Zn, Cd and Mg) compounds computed within DFT are shown in the Fig. 1(a-c). For both ZnV$_{2}$O$_{4}$ and CdV$_{2}$O$_{4}$, the plot of PDOS is divided into I, II, III, IV and V distinct regions, while it is divided into I, II, and III distinct regions for MgV$_{2}$O$_{4}$. For these compounds, V 3$d$ electronic character in the region I (above $\sim$1 eV) of the conduction band (CB) are dominant, where they are more spread for CdV$_{2}$O$_{4}$ as compared to ZnV$_{2}$O$_{4}$ and MgV$_{2}$O$_{4}$. Also in this region, two strong peaks centered around $\sim$2.4 eV and $\sim$2.7 eV for CdV$_{2}$O$_{4}$ and one peak centered around $\sim$2.7 eV for ZnV$_{2}$O$_{4}$ and MgV$_{2}$O$_{4}$ are observed. Similarly, the dominant electronic character in region II comes from the V 3$d$ states, which are roughly extended from $\sim$-1.5 eV of valance band (VB) to $\sim$1 eV of CB for these compounds. In this region, V 3$d$ states centered around Fermi level are less broader for CdV$_{2}$O$_{4}$ as compared to ZnV$_{2}$O$_{4}$ and MgV$_{2}$O$_{4}$. In order to compare the PDOS below $\sim$-3 eV for these compounds, we have multiplied the PDOS for ZnV$_{2}$O$_{4}$, CdV$_{2}$O$_{4}$ and MgV$_{2}$O$_{4}$ by a factor of seven ($\sim$-3 to $\sim$-7 eV), nine ($\sim$-3 to $\sim$-8.5 eV) and five ( below $\sim$-3 eV), respectively. Regions III and IV of VB are extended from $\sim$-3 to $\sim$-7 eV for ZnV$_{2}$O$_{4}$, where V 3$d$ states are mixed with the Zn 3$d$ and O 2$p$ states. These states in regions III and IV are strongly peaked around $\sim$-4.5 eV and $\sim$-6 eV, respectively. Highest electronic character in regions III (IV) comes from the Zn 3$d$ states $\sim$47\% ($\sim$52\%), while lowest comes from V 3$d$ states $\sim$9\% ($\sim$15\%). The regions III and IV of VB for CdV$_{2}$O$_{4}$ and region III of VB for MgV$_{2}$O$_{4}$ are extended from $\sim$-3 to $\sim$-8.5 eV and below $\sim$-3 eV, respectively, where V 3$d$ states are mixed with Cd 4$d$ and O 2$p$ states for CdV$_{2}$O$_{4}$ and O 2$p$ states for MgV$_{2}$O$_{4}$. These states are strongly peaked around $\sim$-4.6 eV and $\sim$-6.3 eV for CdV$_{2}$O$_{4}$ and $\sim$-5.7 eV for MgV$_{2}$O$_{4}$. Among these states, the contribution of O 2$p$ character $\sim$61\% ($\sim$53\%) in regions III (IV) are highest, whereas V 3$d$ character $\sim$10\% ($\sim$21\%) are lowest for CdV$_{2}$O$_{4}$. Similarly for MgV$_{2}$O$_{4}$, the dominant electronic character comes from O 2$p$ states ($\sim$77\%) in region III. In region V of VB, the dominating electronic character below $\sim$-7 eV comes from Zn 3$d$ states of ZnV$_{2}$O$_{4}$ and below $\sim$-8.5 eV, Cd 4$d$ states of CdV$_{2}$O$_{4}$. Now in order to compare our results with the experiment, we have also shown the x-ray photoemission spectroscopy measurements data (where integrated background is subtracted from the experimental data) for ZnV$_{2}$O$_{4}$ and CdV$_{2}$O$_{4}$ in Fig.1 (a and b).\cite{Takubo} For MgV$_{2}$O$_{4}$, we could not find any experimental data available in the literature. It is also clear from the figure that the position of theoretically predicted two peaks in regions III and IV for both compounds are located at $\sim$0.6 eV higher energy than experimentally observed peaks. Normally theoretically predicted peak of O 2$p$ character within DFT is found close to the experiment as shown in the vanadium based compounds.\cite{Haule} 
    
 \begin{figure}[h]
  \begin{center}
    \includegraphics[width=0.45\textwidth]{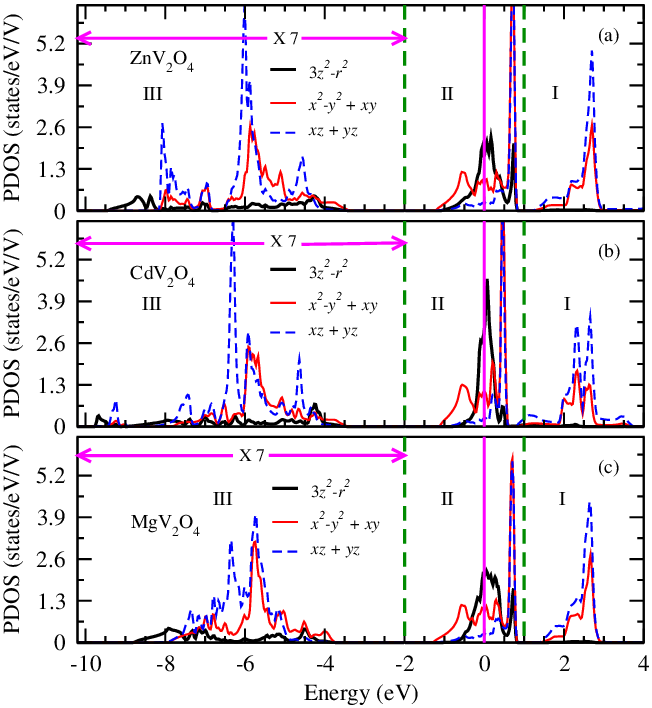}
    \label{}
    \caption{(Color online) The orbitally resolved partial density of states (PDOS) of V 3$d$ states calculated within DFT for (a) ZnV$_{2}$O$_{4}$, (b) CdV$_{2}$O$_{4}$ and (c) MgV$_{2}$O$_{4}$ compounds. For better comparison, PDOS of 3$z^{2}$-$r^{2}$, $x^{2}$-$y^{2}$ + $xy$ and $xz$ + $yz$ orbitals are multiplied by a factor of seven in region III (below$\sim$-2 eV) for these compounds. Zero energy corresponds to the Fermi level.}
    \vspace{-.8cm}
  \end{center}
\end{figure}

    In most of the transition metal oxides, the five fold degenerate 3$d$ orbitals in regular octahedral environment are split into lower energy three fold degenerate t$_{2g}$ and higher energy two fold degenerate e$_{g}$ part, this is not the case here in vanadium spinels. Because of the presence of small trigonal distortion, the degenerate 3$d$ orbitals of V ion are split into non-degenerate 3$z^{2}$-$r^{2}$, doubly degenerate $x^{2}$-$y^{2}$ and $xy$ (indicated by $x^{2}$-$y^{2}$+$xy$) and doubly degenerate $xz$ and $yz$ (indicated by $xz$+$yz$) orbitals. Here, in order know the nature of splitting of these orbitals, we have plotted orbitally resolved PDOS of V atom for these compounds in Fig. 2(a-c). For better comparison, we have multiplied the PDOS of these orbitals by a factor of seven below $\sim$-2 eV for these compounds. The plot of PDOS is divided into I, II and III distinct regions. Above $\sim$1 eV, almost negligible contribution comes to PDOS from 3$z^{2}$-$r^{2}$ orbital for these compounds in region I of CB. In this region, the behaviour of $x^{2}$-$y^{2}$+$xy$ and $xz$+$yz$ orbitals are almost same for these compounds. These orbitals are sharply peaked around $\sim$2.3 eV and $\sim$2.6 eV for CdV$_{2}$O$_{4}$ and $\sim$2.6 eV for ZnV$_{2}$O$_{4}$ and MgV$_{2}$O$_{4}$. $x^{2}$-$y^{2}$+$xy$ and $xz$+$yz$ orbitals are less extended for MgV$_{2}$O$_{4}$ as compared to other two compounds. In region II ($\sim$-2 to $\sim$1 eV) of VB and CB, there is a finite weight of all these orbitals around Fermi level for these spinels, where among these orbitals largest peak of 3$z^{2}$-$r^{2}$ orbital is observed at the Fermi level. However, $x^{2}$-$y^{2}$+$xy$ and $xz$+$yz$ orbitals are sharply peaked as compared to 3$z^{2}$-$r^{2}$ orbital around $\sim$0.5 eV for CdV$_{2}$O$_{4}$ and $\sim$0.7 eV for ZnV$_{2}$O$_{4}$ and MgV$_{2}$O$_{4}$. Around these energies, almost equal contribution from $x^{2}$-$y^{2}$+$xy$ and $xz$+$yz$ orbitals are observed for these spinels. The regions I and II are identified as anti-bonding V 3$d$ orbitals, where the energy of $x^{2}$-$y^{2}$+$xy$ orbitals are lowest and $xz$+$yz$ orbitals are largest as compared to 3$z^{2}$-$r^{2}$ orbital for these spinels. In region III (below $\sim$-2 eV) of VB, V 3$d$ orbitals are extended in largest energy range for CdV$_{2}$O$_{4}$ and smallest for MgV$_{2}$O$_{4}$ as compared to ZnV$_{2}$O$_{4}$. In this region, as compared to other orbitals the contribution from 3$z^{2}$-$r^{2}$ orbital is very small to the PDOS. $xz$+$yz$ orbitals are strongly mixed with $x^{2}$-$y^{2}$+$xy$ orbitals around $\sim$-5.7 eV for these spinels. Also, in this region, $xz$+$yz$ orbitals are sharply peaked around $\sim$-4.5 eV and $\sim$-8.0 eV for ZnV$_{2}$O$_{4}$, $\sim$-4.5 eV and $\sim$-6.4 eV for CdV$_{2}$O$_{4}$ and $\sim$-6.4 eV for MgV$_{2}$O$_{4}$, where the contribution to PDOS from other orbitals are very small. This region is attributed to the bonding V 3$d$ orbitals, where the energy of 3$z^{2}$-$r^{2}$ orbital is lowest and $x^{2}$-$y^{2}$+$xy$ orbitals are largest as compared to $xz$+$yz$ orbitals. The electron occupancy of 3$z^{2}$-$r^{2}$/$xz$+$yz$ ($x^{2}$-$y^{2}$+$xy$) orbitals changes slightly from $\sim$0.65/$\sim$0.71 ($\sim$1.11) to $\sim$0.66/$\sim$0.67 ($\sim$1.13) to $\sim$0.64/$\sim$0.70 ($\sim$1.12) as compound changes from ZnV$_{2}$O$_{4}$ to CdV$_{2}$O$_{4}$ to MgV$_{2}$O$_{4}$. 
    
    In Fig. 3(a-c), we have plotted the total densities of states (TDOS) calculated in DFT+DMFT approach. In order to compare the TDOS with the experimental data (where the total energy resolution was $\sim$0.6 eV), we have included the instrument broadening (IB) of 0.6 eV for these spinels. The TDOS computed by including IB are also shown in the figure. It is clear from the figure that an incoherent shoulder (so called lower Hubbard band) appears at $\sim$-0.9 eV, $\sim$-1.1 eV and $\sim$-1.0 eV for ZnV$_{2}$O$_{4}$, CdV$_{2}$O$_{4}$ and MgV$_{2}$O$_{4}$, respectively, which are slightly mismatch from experimental data. However, by including the IB to TDOS of these compounds, the position of the lower Hubbard band is shifted by $\sim$0.15 eV towards the lower energy side, which is in good agreement with experimental data for ZnV$_{2}$O$_{4}$. However, it is still $\sim$0.2 eV mismatched from the experimental data for CdV$_{2}$O$_{4}$. The behaviour of calculated TDOS is in good agreement with the experiment from Fermi level to $\sim$-4.5 eV for both compounds. However, below $\sim$-4.5 eV, the position of the two peaks is not matched with the experimental data, where similar to DFT data both peaks are located at $\sim$0.6 eV higher energy than experimental results. Now in order to match the position of lower Hubbard band with the experimental data for CdV$_{2}$O$_{4}$, we have increased the values of $U$. For $U$=8 eV and $J$=1.1 eV, the position of the lower Hubbard band is matched with the experimental data as also shown in the figure. The slight mismatching of theoretically calculated lower Hubbard band with the experiment data corresponding to self-consistent calculated parameters for CdV$_{2}$O$_{4}$ may be due to some other reasons, where careful study is needed in this direction.
        
 \begin{figure}[h]
  \begin{center}
    \includegraphics[width=0.45\textwidth]{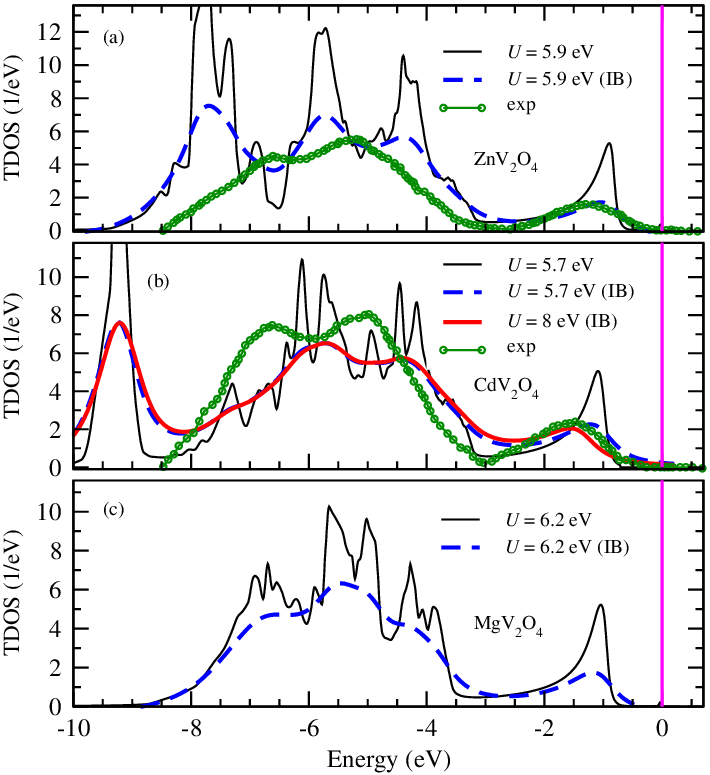}
    \label{}
    \caption{(Color online) Total density of states (TDOS) calculated within DFT+DMFT calculations for (a) ZnV$_{2}$O$_{4}$, (b) CdV$_{2}$O$_{4}$ and (c) MgV$_{2}$O$_{4}$ compounds. TDOS computed by including instrument broadening (IB) are also shown for these compounds. Integrated background subtracted X-ray photoemission spectroscopy measurements data for ZnV$_{2}$O$_{4}$ and CdV$_{2}$O$_{4}$ are taken from the Ref. 41. Zero energy corresponds to the Fermi level.}
    \vspace{-0.5cm}
  \end{center}
\end{figure}

      The plot of PDOS of Zn 3$d$, Cd 4$d$, V 3$d$ ($xz$/$xy$ and $xy$ orbitals) and O 2$p$ states of these spinels computed within DFT+DMFT approach are shown in the Fig. 4(a-f). For ZnV$_{2}$O$_{4}$, the plot of PDOS of Zn 3$d$, V 3$d$ and O 2$p$ states are divided into three distinct regions, I (above Fermi level), II (Fermi level to $\sim$-7 eV) and III (below $\sim$-7 eV). Similarly, the plot of PDOS of Cd 4$d$, V 3$d$ and O 2$p$ states of CdV$_{2}$O$_{4}$ are also divided into three distinct regions, I (above Fermi level), II (Fermi level to $\sim$-8.5 eV) and III (below $\sim$-8.5 eV). While it is divided into two distinct regions, I (above Fermi level) and II (below Fermi level) for MgV$_{2}$O$_{4}$. Now, in order to compare the PDOS for these states, we have multiplied the PDOS of region II by a factor of four for these compounds. It is clear from the figure that the quasiparticle peaks of V 3$d$ character crosses the Fermi level within DFT are disappeared in DFT+DMFT calculations. Hence, the band gap ($E_{g}$) is created for these vanadates. We will discuss the $E_{g}$ for these compounds in the later part of the manuscript in more detail. In regions I and II, two peaks of dominant V 3$d$ character for ZnV$_{2}$O$_{4}$ (CdV$_{2}$O$_{4}$) are sharply peaked around $\sim$3 eV ($\sim$2.7 eV) and $\sim$-0.9 eV ($\sim$-1.1 eV), respectively. Similarly, they are sharply peaked around $\sim$2.9 eV (region I) and $\sim$-1 eV (region II) for MgV$_{2}$O$_{4}$ compound. In region II, Zn 3$d$, Cd 4$d$ and O 2$p$ states are observed below $\sim$-3 eV for these spinels, where they are strongly mixed with V 3$d$ states. The position of the sharp peaks observed here are not shifted appreciably from the DFT results for these compounds. In region III, below $\sim$-7 eV for ZnV$_{2}$O$_{4}$ and below $\sim$-8.5 eV for CdV$_{2}$O$_{4}$, the dominant electron character comes from Zn 3$d$ and Cd 4$d$ states, respectively. The position of these states for both compounds are also not shifted appreciably from the DFT calculations.\\

\begin{figure}[h]
  \begin{center}
    \includegraphics[width=0.45\textwidth]{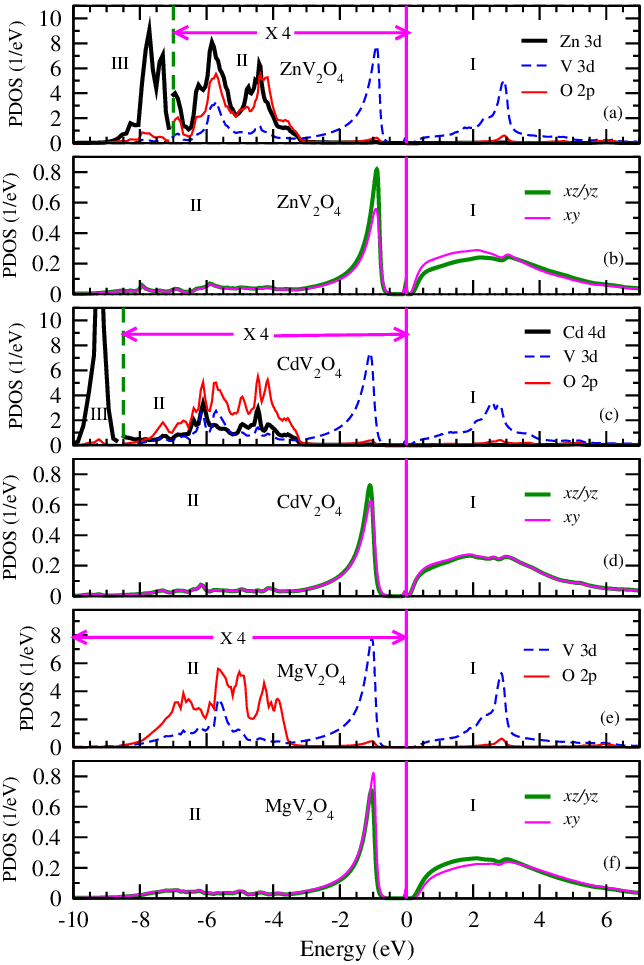}
    \label{}
    \caption{(Color online) Partial density of states (PDOS) of (a) Zn 3$d$, V 3$d$ and O 2$p$ states and (b) $xz$/$yz$ and $xy$ orbitals of V atom for ZnV$_{2}$O$_{4}$, (c) Cd 4$d$, V 3$d$ and O 2$p$ states and (d) $xz$/$yz$ and $xy$ orbitals of V atom for CdV$_{2}$O$_{4}$ and (e) V 3$d$ and O 2$p$ states and (f) $xz$/$yz$ and $xy$ orbitals of V atom for MgV$_{2}$O$_{4}$ compounds computed within DFT+DMFT calculations. For better comparison, PDOS of Zn 3$d$, Cd 4$d$, V 3$d$ and O 2$p$ states in regions II are multiplied by a factor of four for these spinels. Zero energy corresponds to the Fermi level.}
    \vspace{-.8cm}
  \end{center}
\end{figure}

    The plot of PDOS of $xy$ and $xz$/$yz$ orbitals are also shown in the Fig. 4, where it is divided into two distinct regions, I (above Fermi level) and II (below Fermi level). In region I of CB, the incoherent shoulder of these orbitals are observed around $\sim$2 eV for these compounds. In this region, among these orbitals, the large electronic character comes from $xy$ orbital for ZnV$_{2}$O$_{4}$ and CdV$_{2}$O$_{4}$, while it comes from $xz$/$yz$ orbitals for MgV$_{2}$O$_{4}$. The energy of $xy$ orbital is lower than $xz$/$yz$ orbitals for ZnV$_{2}$O$_{4}$ and CdV$_{2}$O$_{4}$. However for MgV$_{2}$O$_{4}$, the energy of $xz$/$yz$ orbitals are found to lower than $xy$ orbital. In region II of VB, all three orbitals are sharply peaked around $\sim$-0.9 eV, $\sim$-1.1 eV and $\sim$-1 eV for ZnV$_{2}$O$_{4}$, CdV$_{2}$O$_{4}$ and MgV$_{2}$O$_{4}$, respectively. In this region, the opposite behaviour of these orbitals are observed as compared to the region I. The partial occupancy of $xy$ orbital increases slightly from $\sim$0.63 to $\sim$0.65 to $\sim$0.69 as compound changes from ZnV$_{2}$O$_{4}$ to CdV$_{2}$O$_{4}$ to MgV$_{2}$O$_{4}$. While the partial occupancy of $xz$/$yz$ orbitals decreases slightly from $\sim$0.69 to $\sim$0.68 to $\sim$0.66 as compound changes from ZnV$_{2}$O$_{4}$ to CdV$_{2}$O$_{4}$ to MgV$_{2}$O$_{4}$. However, the total occupancy of t$_{2g}$ orbitals is $\sim$2.01 for these spinels. The occupancy of $xy$, $xz$/$yz$ and t$_{2g}$ orbitals are about 1.2, 2 and 1.6 times larger than DFT results, respectively. The different orbital occupancy of these orbitals from compound to compound in both approaches are due to the different V-O and V-V bonds and V-O-V angles in the trigonal distorted VO$_{6}$ octahedra for these spinels. Different bonds and angles influence the orbital occupancy are due to the different ionic radii of Zn, Cd and Mg ions in these compounds.

   The calculated values of $E_{g}$ for ZnV$_{2}$O$_{4}$, CdV$_{2}$O$_{4}$ and MgV$_{2}$O$_{4}$ are $\sim$0.9 eV, $\sim$0.95 eV and $\sim$1.15 eV, respectively. For ZnV$_{2}$O$_{4}$ and MgV$_{2}$O$_{4}$ compounds, the values of $E_{g}$ computed in the present study are about three times larger than the experimental data reported by Rogers $et$ $al$. However, the experimentally observed values of $E_{g}$ by Pardo $et$ $al$. are $\sim$1.1 eV, $\sim$1.08 eV and $\sim$3.16 eV for ZnV$_{2}$O$_{4}$, MgV$_{2}$O$_{4}$ and CdV$_{2}$O$_{4}$, respectively. The calculated values of $E_{g}$ are close to the experimental values for ZnV$_{2}$O$_{4}$ and MgV$_{2}$O$_{4}$ compounds. However, it is about three times smaller than the experimental data for CdV$_{2}$O$_{4}$. As mentioned above for $U$=8 eV and $J$=1.1 eV, the position of the lower Hubbard band match to the experimental result for CdV$_{2}$O$_{4}$ compound. Corresponding to these parameters, DFT+ DMFT approach still underestimate the value of $E_{g}$ ($\sim$1.6 eV) as compared to experimental result ($\sim$3.16 eV) for this compound. Experimentally, the value of $E_{g}$ depends on the activation energy for the resistivity, which itself depends whether the spinel sample is made in pure phase or not. However, it is always a challenging task to make the spinel sample in pure phase. In the light of these facts, it is difficult to compare the theoretically computed and experimentally observed values of $E_{g}$ for these compounds. For these parameters, the values of $E_{g}$ computed in the present study for these compounds are reasonable as corresponding to these parameters, our data is closely matched with the experiment in VB. Here, it will be interesting to see whether the inverse photoemission spectroscopy measurements data will be helpful to explain our results in CB for these compounds.
    
 \begin{figure}[h]
  \begin{center}
    \includegraphics[width=0.45\textwidth]{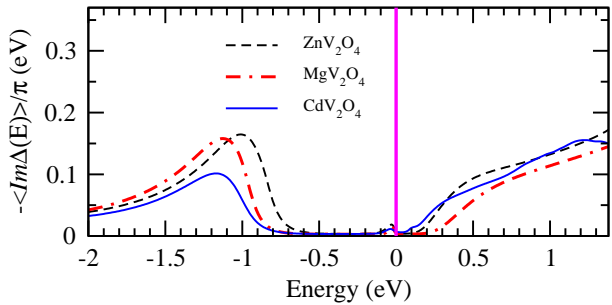}
    \label{}
    \caption{(Color online) The imaginary part of the average impurity hybridization function (average over all t$_{2g}$ orbitals of V site computed within DFT+DMFT calculations for AV$_{2}$O$_{4}$ (A=Zn, Cd and Mg) compounds. Zero energy corresponds to the Fermi level.}
    \vspace{-.9cm}
  \end{center}
\end{figure}

    It is well known that in insulating vanadium spinels, the localization of V 3$d$ electrons are not in fully localized limit. Based on localized-electron superexchange represented by $J$$\propto$$\frac{t^{2}}{U}$ (where, $t$ is the transfer integral between neighbouring sites), Canosa $et$ $al$. have argued qualitatively that the order of degree of localization of these electrons for CdV$_{2}$O$_{4}$$>$ZnV$_{2}$O$_{4}$$>$MgV$_{2}$O$_{4}$.\cite{Canosa} However, based on the calculated values of ratio between average values of nearest neighbour $t$ and self-consistently computed values of $U$ in our earlier work, we have shown that the degree of localization of these electrons decreases from CdV$_{2}$O$_{4}$ to MgV$_{2}$O$_{4}$ to ZnV$_{2}$O$_{4}$.\cite{Lalarxiv} In that study, the average values of $t$ for these spinels were computed indirectly by fitting the following equation,\cite{Saul}

\begin{align*}    
    J\approx-\frac{4t^{2}}{U} 
\end{align*}   

valid to localized limit. $J$ was the average nearest neighbour exchange coupling constant. Here, based on the impurity hybridization function (gives the direct information about the degree of localization of correlated electrons), it is interesting to see the order of degree of localization of V 3$d$ electrons in these spinels. Hence, we have computed the average impurity hybridization function (average over all t$_{2g}$ orbitals of V site) for these compounds, which is shown in the Fig. 5. Below Fermi level, the average impurity hybridization function is sharply peaked around $\sim$-1 eV, $\sim$-1.1 eV and $\sim$-1.15 eV for ZnV$_{2}$O$_{4}$, MgV$_{2}$O$_{4}$ and CdV$_{2}$O$_{4}$, respectively. Around these energies, it is largest for ZnV$_{2}$O$_{4}$ and smallest for CdV$_{2}$O$_{4}$ as compared to MgV$_{2}$O$_{4}$. Large value of the average impurity hybridization function indicates the less localized electronic character for the system. Hence, the degree of localization of V 3$d$ electrons decreases from CdV$_{2}$O$_{4}$ to MgV$_{2}$O$_{4}$ to ZnV$_{2}$O$_{4}$, which is consistent with our earlier results based on above discussed $\frac{t}{U}$ ratio.

\section{Conclusions} 
    
    In conclusion, the exact comparative behaviour of various physical properties of AV$_{2}$O$_{4}$ (A=Zn, Cd and Mg) compounds have not been found, where material specific parameters were used. In present work, we have explored the comparative behaviour of electronic structures of these compounds by using DFT+DMFT approach, where material specific parameters were used. The calculated order of $E_{g}$ was found to be MgV$_{2}$O$_{4}$ ($\sim$1.15 eV)$>$CdV$_{2}$O$_{4}$ ($\sim$0.95 eV)$>$ZnV$_{2}$O$_{4}$ ($\sim$0.9 eV), where the values of $E_{g}$ were close to experimental data for ZnV$_{2}$O$_{4}$ and MgV$_{2}$O$_{4}$. Among these spinels, the average impurity hybridization function of V site was found to be largest for ZnV$_{2}$O$_{4}$ and smallest for CdV$_{2}$O$_{4}$. Hence, the degree of localization of V 3$d$ electrons was highest for CdV$_{2}$O$_{4}$ and lowest for ZnV$_{2}$O$_{4}$ as compared to MgV$_{2}$O$_{4}$ compound. This order of degree of localization of V 3$d$ electrons was found to be consistent with our earlier results. In the energy range of 0 to $\sim$-4.5 eV, almost a similar behaviour of calculated total density of states and experimental data were observed for these spinels, where the position of lower Hubbard band was in good agreement with the experiment for ZnV$_{2}$O$_{4}$.
 
\section{Acknowledgements}

  S.L. is thankful to UGC, India, for financial support.


\begin{thebibliography}{99}

\bibitem{Nishiguchi} N. Nishiguchi, and M. Onoda, J. Phys: Condens. Matter {\bf 14}, L551 (2002).

\bibitem{Reehuis} M. Reehuis, A. Krimmel, N. B$\ddot{\rm u}$ttgen, A. Loidl, and A. Prokofiev, Eur. Phys. J. B {\bf 35}, 311 (2003).

\bibitem{Tsunetsugu} H. Tsunetsugu, and Y. Motome, Phys. Rev. B {\bf 68}, 060405(R) (2003). 

\bibitem{Lee} S. -H. Lee, D. Louca, H. Ueda, S. Park, T. J. Sato, M. Isobe, Y. Ueda, S. Rosenkranz, P. Zschack, J. $\acute{\rm I}$$\tilde{\rm n}$iguez, Y. Qiu, and R. Osborn, Phys. Rev. Lett. {\bf 93}, 156407 (2004).

\bibitem{Tchernyshyov} O. Tchernyshyov, Phys. Rev. Lett. {\bf 93}, 157206 (2004).

\bibitem{Matteo} S. Di Matteo, G. Jackeli, and N. B. Perkins, Phys. Rev. B {\bf 72}, 020408(R) (2005). 

\bibitem{Motome} H. Tsunetsugu, and Y. Motome, Prog. Theor. Phys. Suppl. {\bf 160}, 203 (2005).

\bibitem{Suzuki} T. Suzuki, M. Katsumura, K. Taniguchi, T. Arima, and T. Katsufuji, Phys. Rev. Lett. {\bf 98}, 127203 (2007).

\bibitem{Maitra} T. Maitra, and R. Valent$\acute{\rm \oldstylenums{1}}$, Phys. Rev. Lett. {\bf 99}, 126401 (2007).

\bibitem{Wheeler} E. M. Wheeler, B. Lake, A. T. M. Nazmul Islam, M. Reehuis, P. Steffens, T. Guidi, and A. H. Hill, Phys. Rev. B {\bf 82}, 140406(R) (2010).

\bibitem{Pandey2011} S. K. Pandey, Phys. Rev. B {\bf 84}, 094407 (2011); Phys. Rev. B {\bf 86}, 085103 (2012).

\bibitem{Lal1} S. Lal, and S. K. Pandey, Eur. Phys. J. B {\bf 87}, 197 (2014); J. Magn. Magn. Mater. {\bf 412}, 23 (2016); Comput. Mater. Sci. {\bf 126}, 373 (2017).

\bibitem{Lal3} S. Lal, and S. K. Pandey, Mater. Res. Exp. {\bf 3}, 116301 (2016).

\bibitem{Lal5} S. Lal, and S. K. Pandey,  Phys. Lett. A {\bf 381}, 917 (2017).

\bibitem{Takagi} H. Takagi, and S. Niitaka, {\it Introduction to Frustrated Magnetism}, Springer Series in Solid-State Sciences, Vol. 164 (Springer, Berlin) 2011, Part 3, p. 155 and references therein.

\bibitem{Rogers} D.B. Rogers, R.J. Arnott, A. Wold, J.B. Goodenough, J. Phys. Chem. Solids {\bf 24}, 347 (1963).

\bibitem{Pardo} V. Pardo, S. Blanco-Canosa, F. Rivadulla, D. I. Khomskii, D. Baldomir, H. Wu, and J. Rivas, Phys. Rev. Lett. {\bf 101}, 256403 (2008).

\bibitem{Canosa} S. Blanco-Canosa, F. Rivadulla, V. Pardo, D. Baldomir, J.-S. Zhou, M. Garcia-Hern$\acute{a}$ndez, M. A. Lopez-Quintela, J. Rivas, and J.B. Goodenough, Phys. Rev. Lett. {\bf 99}, 187201 (2007).

\bibitem{Lalarxiv} S. Lal, and S. K. Pandey, {\it arXiv:1611.02028v1}.

\bibitem{Yamashita} Y. Yamashita, and K. Ueda, Phys. Rev. Lett. {\bf 85}, 4960 (2000).

\bibitem{Khomskii} D. I. Khomskii, and T. Mizokawa, Phys. Rev. Lett. {\bf 94}, 156402 (2005).

\bibitem{Kaur} R. Kaur, T. Maitra, and T. Nautiyal, J. Phys: Condens. Matter {\bf 25}, 065503 (2013).

\bibitem{Zhu} J. -X. Zhu, R. C. Albers, K. Haule, G. Kotliar, and J. M. Wills, Nat. Commun. {\bf 4}, 2644 (2013), and references therein. 

\bibitem{Georges1992} A. Georges, and G. Kotliar, Phys. Rev. B {\bf 45}, 6479 (1992).  

\bibitem{Georges1996} A. Georges, G. Kotliar, W. Krauth, and M. J. Rozenberg, Rev. Mod. Phys. {\bf 68}, 13 (1996).

\bibitem{Anisimov} V. I. Anisimov, A. I. Poteryaev, M. A. Korotin A. O. Anokhin, and G. Kotliar, J. Phys.: Condens. Matter {\bf 9}, 7359 (1997).

\bibitem{Kotliar} G. Kotliar, S. Y. Savrasov, K. Haule, V. S. Oudovenko, O. Parcollet, and C. A. Marianetti, Rev. Mod. Phys. {\bf 78}, 865 (2006). 

\bibitem{Martins} C. Martins, M. Aichhorn, L. Vaugier, and S. Biermann, Phys. Rev. Lett. {\bf 107}, 266404 (2011).

\bibitem{Biermann} S. Biermann, {\it Encyclopedia of Materials: Science and Technology} (Elsevier Ltd., New York, 2006). 

\bibitem{Roekeghem} A. V. Roekeghem, and S. Biermann, Europhys. Lett. {\bf 108}, 57003 (2014).

\bibitem{Haule} K. Haule, T. Birol, and G. Kotliar, Phys. Rev. B {\bf 90}, 075136 (2014), and references therein.

\bibitem{Wang} X. Wang, M. J. Han, L. de Medici, H. Park, C. A. Marianetti, and A.J. Millis, Phys. Rev. B {\bf 86}, 195136 (2012).

\bibitem{Dang} H. T. Dang, A. J. Millis, and C. A. Marianetti, Phys. Rev. B {\bf 89}, 161113(R) (2014).

\bibitem{Blaha} P. Blaha, K. Schwarz, G. K. H. Madsen, D. Kvasnicka, and J. Luitz, WIEN2k, An Augmented Plane Wave Plus Local Orbitals Program for Calculating Crystal Properties (Vienna University of Technology, Vienna, 2001).

\bibitem{Perdew} J. P. Perdew, A. Ruzsinszky, G. I. Csonka, O. A. Vydrov, G. E. Scuseria, L. A. Constantin, X. Zhou, and K. Burke, Phys. Rev. Lett. {\bf 100}, 136406 (2008).

\bibitem{Onoda} M. Onoda, and J. Hasegawa, J. Phys.: Condens. Matter {\bf 15}, L95 (2003).

\bibitem{Haule2010} K. Haule, C.-H. Yee, and K. Kim, Phys. Rev. B {\bf 81}, 195107 (2010).

\bibitem{Haule2007} K. Haule, Phys. Rev. B {\bf 75}, 155113 (2007).

\bibitem{Haule2015} K. Haule,  Phys. Rev. Lett. {\bf 115}, 196403 (2015).

\bibitem{Jarrell} M. Jarrell, and J. E. Gubernatis, Phys. Rep. {\bf 269}, 133 (1996). 

\bibitem{Takubo} K. Takubo, J.-Y. Son, T. Mizokawa, H. Ueda, M. Isobe, Y. Matsushita, and Y. Ueda Phys. Rev. B {\bf 74}, 155103 (2006). 

\bibitem{Saul} A. Saul, and G. Radtke, Phys. Rev. Lett. {\bf 106}, 177203 (2011).

\end{thebibliography}
\end{document}